\documentclass[review]{elsarticle}

\usepackage{lipsum}
\usepackage{hyperref}
\usepackage{inputenc}
\usepackage{graphicx}
\usepackage{amsmath}
\usepackage[official]{eurosym}

\journal{Medical Engineering and Physics}

\begin{document}

\begin{frontmatter}

\title{Dynamic simulation of aortic valve stenosis using a lumped parameter cardiovascular system model with flow regime dependent valve pressure loss characteristics}

\author{Ryno Laubscher}
\address{Institute for Biomedical Engineering, Department of Mechanical and Mechatronic Engineering, Stellenbosch University, Stellenbosch, South Africa}
\author{Jacques Liebenberg and Philip Herbst}
\address{Division of Cardiology, Faculty of Medicine and Health Sciences, Stellenbosch University, Cape Town, South Africa}

\begin{abstract}

Valvular heart diseases are growing concern in impoverished parts of the world, such as Southern-Africa, claiming more than 31 \% of total deaths related to cardiovascular diseases. The ability to model the effects of regurgitant and obstructive lesions on the valve body can assist clinicians in preparing personalised treatments. In the present work, a multi-compartment lumped parameter model of the human cardiovascular system is developed, with a newly proposed valve modelling approach which accounts for geometry and flow regime dependent pressure drops along with the valve cusp motion. The model is applied to study various degrees of aortic stenosis using typical human cardiovascular parameters. The results generated with the proposed model, are compared to predictions using previously published valve modelling approaches and both sets of results are compared to typical local and global physiological parameters found in literature such left-ventricular systolic pressures, peak and mean aortic valve pressure drops and vena contracta velocities. The results show that the previously published valve models under predicts expected severely stenosed peak and mean transvalvular pressure drops by approximately 47\% and 30\% respectively, whereas the newly proposed model under predicts the peak pressure drop by 20\% and over predicts mean pressure drop by 7\%.

\end{abstract}


\end{frontmatter}


\section{Introduction}

Cardiovascular diseases (CVDs) are a growing social and healthcare burden on the world. Recent studies show that globally an estimated 19 million people died \cite{Tsao2022} from CVDs in 2020 compared to 17.9 million in 2017 \cite{WorldHealthOrgani2021}. The leading cause of CVD related deaths in high-income countries such as in Europe and Northern America is ischaemic heart disease (IHD), which accounts for approximately 50\% of deaths from CVDs. In impoverished countries, such as those located in Africa, IHD accounts for less than 10\% of deaths from CVDs. In Sub-Saharan Africa, valvular heart diseases such as rheumatic heart disease (RHD), bicuspid aortic valve disease and degenerative aortic valve disease, accounts for approximately 31\% of CVD related deaths \cite{Yuyun2020},\cite{Keates2017}. Recent statistical projections also show that valvular diseases are set to double in the United States and Europe before 2050 \cite{Tsao2022}. 

Valvular heart disease is broadly categorised as obstructive and regurgitant lesions of the valve body. Valve malfunction, whether due to regurgitation or obstruction is associated with abnormal intracardiac haemodynamic behaviour. Left untreated this can lead to reduced cardiac output or high ventricular pressures \cite{Rubenstein2022}. The ability to quantify the influence of valvular diseases on the local and global haemodynamics of the cardiovascular system would assist clinicians in evaluating patient specific cardiovascular performance and preparing personalised treatments \cite{Garber2021}. One approach to non-invasively predict and study valve disease effects on local and global haemodynamic functions is to use electrical-hydraulic analogue lumped-parameter mathematical models (LPMs) of the cardiovascular system \cite{Hose2019}. LPMs can be used to quickly simulate variations in blood pressures, flow rates and volumes of various compartments that make up the circulatory system as a function of time.

Recently several authors have utilised LPMs to study various CVDs \cite{Rosalia2021}, \cite{Mao2019}, \cite{Bozkurt2019}, \cite{Shimizu2018}, \cite{Keshavarz-Motamed2020}, \cite{Tang2020}. Garber et al. \cite{Garber2021} stated that the heart valve modelling approaches predominantly used in literature, are merely simplifications of the actual valve mechanics and does not take into account the local haemodynamics such as mechanical forces and valve motion. Some researchers have developed more advanced valve models for LPMs, which included valve motion and area-dependent pressure losses. Korakianitis and Shi \cite{Korakianitis2006a}, \cite{Korakianitis2006} studied the effect of aortic regurgitation and mitral stenosis  using a LPM and an advanced valve model which accounted for valve motion and variable area effects on pressure losses. Similarly, Mynard et al. \cite{Mynard2011} proposed an advanced valve model which incorporates the valve motion as a function of the pressure forces. The proposed model used tuned coefficients to accurately capture the valve motion and pressure losses through the valves.

In the present work, a 0D valve model is proposed which accounts for valve motion and Reynolds number dependent pressure losses. The model, similarly to \cite{Korakianitis2006}, solves the valve cusp dynamics as a set of differential algebraic equations (DAEs) and uses the valve position to calculate the pressure losses. To calculate the pressure drop through the valve, the domain is approximated as a gradual contraction (nozzle) rather than a typical orifice and includes the entrance, exit, frictional and local loss effects. The proposed model is applied to investigate the influence of increasing degrees of aortic stenosis on a cardiovascular system using a multi-compartment LPM. To investigate if the proposed level of valve modelling detail is necessary, the model outputs are compared to results generated using the approaches presented by \cite{Korakianitis2006}. According to the best knowledge of the authors the present work is the first instance of using a dynamic valve model with Reynolds number dependent pressure losses to study the local and global effects of aortic stenosis on a LPM of the circulatory system.

The computer models of the cardiovascular system, were developed using the \emph{Julia} v.1.7.0 \cite{Julia2022} programming language and the \emph{DifferentialEquations.jl} \cite{Rackauckas2017} and \emph{NLSolve.jl} open-source libraries.

\section{Material and methods}

\begin{figure}[h!]
	\centering
	\includegraphics[width=\textwidth]{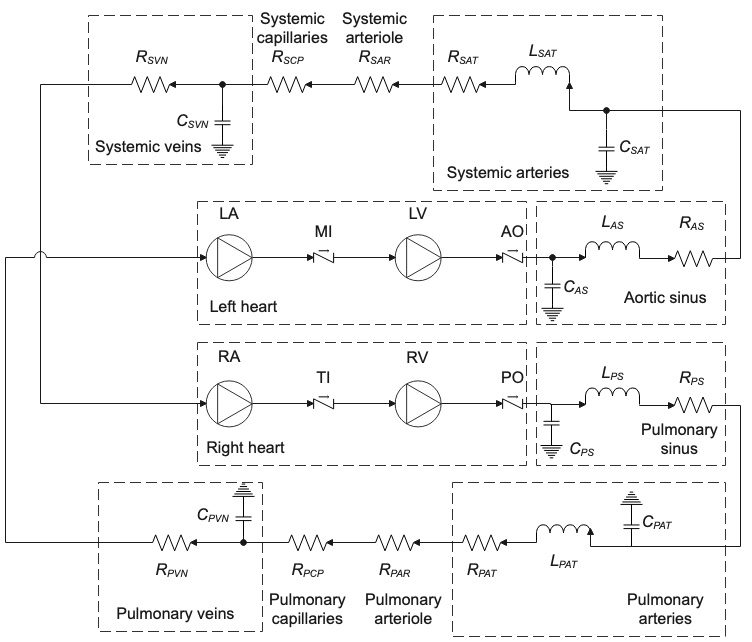}
	\caption{Lumped parameter network model of cardiovascular system. aortic - AO, mitral - MI, pulmonary - PO, tricuspid - TI, aortic sinus - AS, systemic arteries - SAT, systemic arterioles - SAR, systemic capillaries - SCP, systemic vein - SVN, pulmonary sinus - PS, pulmonary arteries - PAT, pulmonary arteriole - PAR, pulmonary capillaries - PCP, pulmonary veins - PVN}
	\label{lpm_circuit}
\end{figure}

In the current section, the mathematical models for the different components in the cardiovascular system seen in figure \ref{lpm_circuit} will be discussed. The LPM developed in the present work consists of the four heart chambers and their respective downstream valves, the systemic circulation network and the pulmonary circulation network.

\subsection{Modelling of heart chambers}\label{heart_section}

The heart chambers act as non-linear pressure sources in the hydraulic network, which pressurises the low-pressure blood coming from the upstream veins or chambers. To simulate the pressure-volume-time relationships of the left ventricle (LV), left atrium (LA), right ventricle (RV) and right atrium (RA) the time-varying elastance model of Suga et al. \cite{Suga1973} is used. The instantaneous static pressure inside the $i^{th}$ chamber (ventricles or atriums) is calculated using the relation shown in equation \ref{heart_pressure}. The instantaneous volume $V_i(t)$ $\left[ mL \right]$ of the $i^{th}$ chamber is calculated using a simple mass balance over the respective heart chamber, as shown in equation \ref{heart_vol}.

\begin{equation}
	P_i(t) = P_{i,0} + e_{i}(t) \left( V_i(t) - V_{i,0} \right)
	\label{heart_pressure}
\end{equation}

\begin{equation}
	\frac{dV_i}{dt} = Q_{i,in}(t) - Q_{i,ex}(t)
	\label{heart_vol}
\end{equation}

In equations \ref{heart_pressure}, \ref{heart_vol}: $P_{i,0}$ $\left[mmHg \right]$ is the unstressed chamber internal pressure, $V_{i,0}$ $\left[ mL \right]$ is the unstressed chamber volume, $Q_{i,in}(t)$ $\left[ mL/s \right]$ is the inlet flow rate into the chamber and $Q_{i,ex}(t)$ the exit flow rate out of the chamber. $e_{i}(t)$ is the time-varying elastance of the $i^{th}$ chamber and is estimated differently for the atriums and ventricles. For the ventricles the time-varying elastances are modelled using equation \ref{heart_ventricle_elastance} \cite{Korakianitis2006} and the ventricular activation functions with equation \ref{heart_act_func_v} \cite{Bozkurt2019}.

\begin{equation}
	e_{i}(t) = E_{i,d} + \frac{E_{i,s} - E_{i,d}}{2}\cdot f_i(t)
	\label{heart_ventricle_elastance}
\end{equation}

\begin{equation}
	f_i(t) = 
	\begin{cases}
		1 - \cos \left[ \left( \frac{t}{T_1} \pi \right) \right], & \text{if } 0 \leq t < T_1 \\
		1 + \cos \left[ \frac{t - T_1}{T_2 - T_1} \pi \right], & \text{if } T_1 \leq t < T_2 \\
		0, & \text{if } T_2 \leq t < T
		
	\end{cases}
	\label{heart_act_func_v}
\end{equation}

In equation \ref{heart_act_func_v}, $T$ $\left[ s \right]$ is the heart beat period, $T_1$ is the time at end of systole and $T_2$ is the time at end of ventricular relaxation. The time-varying elastances for the atriums are calculated using equation \ref{heart_atrium_elastance} and the atrial activation functions are estimated using equation \ref{heart_act_func_a}.

\begin{equation}
	e_i(t) = E_{i,min} + \frac{E_{i,max} - E_{i,min}}{2}\cdot f_i(t - D)
	\label{heart_atrium_elastance}
\end{equation}

\begin{equation}
	f_i(t) =
	\begin{cases}
		0, & \text{if } 0 \leq t < T_a \\
		1 - \cos \left[ 2\pi \frac{t - T_a}{T - T_a} \right], & \text{if } T_a \leq t < T 
	\end{cases}
	\label{heart_act_func_a}
\end{equation}

In equations \ref{heart_atrium_elastance}, \ref{heart_act_func_a}: $E_{i,max}$, $E_{i,min}$ are the maximum and minimum pressure-volume relations inside the atriums, $T_a$ is the time at onset of atrial contraction and $D$ is the time of atrial relaxation.
As seen above, in the time-varying elastance model equations, various empirical values are required, such as $T_1$, $T_2$ and $E_{i,s}$. Table \ref{heart_table_a} contains the empirical values used for the ventricles and atriums. In the current work the heart beat period is fixed at $T=1$ $\left[ s \right]$.

\begin{table}[h!]
	\centering
	\caption{Time-varying elastance model parameters for atriums. Ref. \cite{Bozkurt2019}, \cite{Korakianitis2006a}}
	\begin{tabular} { c c c }
		\hline
		Parameters & Left heart & Right heart \\
		\hline
		&\emph{Atriums} \\
		$E_{max}$ $\left[ \frac{mmHg}{mL} \right]$ & $0.25$ & $0.15$ \\
		$E_{min}$ $\left[ \frac{mmHg}{mL} \right]$ & $0.25$ & $0.15$ \\
		$T_a$ & $0.8 T$ & $0.8 T$ \\
		$D$ $\left[ s \right]$ & $0.04$ & $0.04$ \\
		$V_{0}$ $\left[ mL \right]$ & $4$ & $4$ \\
		$P_{0}$ $\left[ mmHg \right]$ & $1$ & $1$ \\
		&\emph{Ventricles} \\
		$E_{d}$ $\left[ \frac{mmHg}{mL} \right]$ & $0.1$ & $0.1$ \\
		$E_{s}$ $\left[ \frac{mmHg}{mL} \right]$ & $2.5$ & $1.15$ \\
		$T_1$ & $0.3 T$ & $0.3 T$ \\
		$T_2$ & $0.45 T$ & $0.45 T$ \\
		$V_{0}$ $\left[ mL \right]$ & $5$ & $10$ \\
		$P_{0}$ $\left[ mmHg \right]$ & $1$ & $1$ \\
		\hline
	\end{tabular}	
	\label{heart_table_a}
\end{table}

\subsection{Systemic and pulmonary networks}\label{vasculature_section}
The systemic and pulmonary vasculature networks are modelled using five components each, as seen in figure \ref{lpm_circuit}. These components are the sinuses, arteries, arterioles, capillaries and veins \cite{Korakianitis2006a}. Depending on the type of vasculature, the local haemodynamics is simulated using a combination of hydraulic resistance ($R$), blood inertia (inductance, $L$) and vessel compliance (capacitance, $C$) to capture the time-dependent pressure-volume behaviour of the specific vessel. 

The flow rate through the aortic sinus is simulated using the ordinary differential equation (ODE) shown in equation \ref{AS_flow}. The time-dependent blood pressure inside the aortic sinus can be evaluated by solving the flow rate differential equation and equation \ref{AS_P} simultaneously \cite{Naik2017}, \cite{Korakianitis2006a}.

\begin{equation}
	L_{AS} \frac{d Q_{AS}}{dt} = \left( P_{AS} - P_{SAT} \right) - R_{AS} Q_{AS}
	\label{AS_flow}
\end{equation}

\begin{equation}
	C_{AS} \frac{d P_{AS}}{dt} = Q_{AO} - Q_{AS}
	\label{AS_P}
\end{equation}

In the equations above, $L_{AS}$ $\left[ mmHg\cdot s^2/mL \right]$ is the blood inertia at the aortic sinus, $Q_{AS}$ $\left[ mL/s \right]$ is the flow rate of blood out of the sinus, $P_{AS}$ $\left[ mmHg \right]$ is the inlet aortic sinus pressure, $P_{SAT}$ $\left[ mmHg \right]$ is the systemic arterial inlet pressure, $R_{AS}$ $\left[ mmHg \cdot s / mL \right]$ is the aortic sinus hydraulic resistance, $C_{AS}$ $\left[ mL/mmHg \right]$ is the aortic sinus compliance and $Q_{AO}$ is the flow rate of blood leaving the aortic valve. 

For the pulmonary sinus the flow rate and blood pressure can be simulated by solving equations \ref{PS_flow}, \ref{PS_P}. The descriptions and units of the variables are similar to that of the aortic sinus above.

\begin{equation}
	L_{PS} \frac{d Q_{PS}}{dt} = \left( P_{PS} - P_{PAT} \right) - R_{PS} Q_{PS}
	\label{PS_flow}
\end{equation}

\begin{equation}
	C_{PS} \frac{d P_{PS}}{dt} = Q_{PO} - Q_{PS}
	\label{PS_P}
\end{equation}

The systemic arteries are modelled similarly to the aortic sinus. The governing differential equations for flow rate and inlet blood pressure for the systemic arteries is given in equations \ref{SAT_flow}, \ref{SAT_P}.

\begin{equation}
	L_{SAT} \frac{d Q_{SAT}}{dt} = \left( P_{SAT} - P_{SAR} \right) - R_{SAT} Q_{SAT}
	\label{SAT_flow}
\end{equation}

\begin{equation}
	C_{SAT} \frac{d P_{SAT}}{dt} = Q_{AS} - Q_{SAT}
	\label{SAT_P}
\end{equation}

The variations in pulmonary arteries outlet flow rate and inlet blood pressure are simulated by solving equations \ref{PAT_flow}, \ref{PAT_P}.

\begin{equation}
	L_{PAT} \frac{d Q_{PAT}}{dt} = \left( P_{PAT} - P_{PAR} \right) - R_{PAT} Q_{PAT}
	\label{PAT_flow}
\end{equation}

\begin{equation}
	C_{PAT} \frac{d P_{PAT}}{dt} = Q_{PS} - Q_{PAT}
	\label{PAT_P}
\end{equation}

The systemic arteriole and capillaries are modelled using only hydraulic resistance components due to the rigidity and small diameter of the vessel walls \cite{Tang2020}. The flow though these vasculature sections are assumed to be steady, thus $Q_{SAT} = Q_{SAR} = Q_{SCP}$. The pressure drop through these sections are calculated using equations \ref{SAR_P}, \ref{SCP_P}.

\begin{equation}
	R_{SAR} Q_{SAR} = P_{SAR} - P_{SCP} \equiv 	R_{SAR} Q_{SAT} = P_{SAR} - P_{SCP} 
	\label{SAR_P}
\end{equation} 

\begin{equation}
	R_{SCP} Q_{SCP} = P_{SCP} - P_{SVN} \equiv R_{SCP} Q_{SAT} = P_{SCP} - P_{SVN}
	\label{SCP_P}
\end{equation} 

Similarly, for the pulmonary network arteriole and capillaries the pressure drops are calculated using equations \ref{PAR_P}, \ref{PCP_P}. 

\begin{equation}
	R_{PAR} Q_{PAR} = P_{PAR} - P_{PCP} \equiv 	R_{PAR} Q_{PAT} = P_{PAR} - P_{PCP} 
	\label{PAR_P}
\end{equation} 

\begin{equation}
	R_{PCP} Q_{PCP} = P_{PCP} - P_{PVN} \equiv R_{PCP} Q_{PAT} = P_{PCP} - P_{PVN}
	\label{PCP_P}
\end{equation} 

The systemic and pulmonary veins are simulated as compliance-resistance components with negligible blood inertia. The governing differential equations for the systemic (equation \ref{SVN_P}) and pulmonary (equation \ref{PVN_P}) vasculature inlet pressures are shown below.

\begin{equation}
	C_{SVN} \frac{d P_{SVN}}{dt} = Q_{SCP} - Q_{SVN}
	\label{SVN_P}
\end{equation}

\begin{equation}
	C_{PVN} \frac{d P_{PVN}}{dt} = Q_{PCP} - Q_{PVN}
	\label{PVN_P}
\end{equation}

To determine the flow rate through the venous system the hydraulic resistance expressions are used. The driving pressure for the systemic vein flow rate is the pressure difference between the vein inlet and the right atrium chamber pressure \cite{Naik2017}. For the pulmonary network, the driving pressure is between the inlet to the veins and the left atrium chamber pressure. 

\begin{equation}
	Q_{SVN} R_{SVN} = P_{SVN} - P_{RA}
	\label{SVN_Q}
\end{equation}

\begin{equation}
	Q_{PVN} R_{PVN} = P_{PVN} - P_{LA}
	\label{SVN_Q}
\end{equation}

All parameters used in the vasculature model are shown in table \ref{sys_pul_net_pars}. In the next section the modelling of the different valve modelling approaches will be discussed.

\begin{table}[h!]
	\centering
	\caption{Systemic and pulmonary vasculature network parameters}
	\begin{tabular} {c c c c c}
	\hline
		Parameters & $R$ $\left[ \frac{mmHg \cdot s}{mL} \right]$ & $L$ $\left[ \frac{mmHg \cdot s^2}{mL} \right]$ & $C$ $\left[ \frac{mL}{mmHg} \right]$ & Ref. \\
	\hline	
	\emph{Systemic network} \\
	Aortic sinus & 0.003 & $6.2\cdot 10^{-5}$ & 0.08 & \cite{Korakianitis2006a} \\
	Systemic arteries & 0.05 & 0.0017 & 1.6 & \cite{Korakianitis2006a}, \cite{Sun1995} \\
	Systemic arterioles & 0.5 & - & - & \cite{Korakianitis2006a}, \cite{Korakianitis2006} \\
	Systemic capillaries & 0.52 & - & - & \cite{Korakianitis2006a}, \cite{Korakianitis2006} \\
	Systemic veins & 0.075 & - & 22.0 & \cite{Korakianitis2006a}, \cite{Korakianitis2006} \\
	\emph{Pulmonary network} \\
	Pulmonary sinus & 0.002 & $5.2 \cdot 10^{-5}$ & 0.18 & \cite{Korakianitis2006a}, \cite{Korakianitis2006} \\
	Pulmonary arteries & 0.01 & 0.0017 & 5.0 & \cite{Korakianitis2006a},\cite{Bozkurt2019} \\
	Pulmonary arterioles & 0.05 & - & - & \cite{Korakianitis2006a}, \cite{Korakianitis2006} \\
	Pulmonary capillaries & 0.05 & - & - & - \\
	Pulmonary veins & 0.006 & - & 30.0 & \cite{Bozkurt2019}, \cite{Korakianitis2006a} \\
	\hline
	\end{tabular} \\
	\label{sys_pul_net_pars}
\end{table}

\subsection{Heart valve models}\label{heart_valve_section}

The purpose of the four heart valves is to prevent reverse flow or regurgitation of blood back into the heart chambers, similar to the operation of check valves used in the process industry. The opening and closing processes of the valve cusps are governed by a combination of pressure, vortex and frictional forces \cite{Korakianitis2006a} acting on the surfaces of the valves. In typical lumped parameter modelling of cardiovascular systems, heart valves are approximated as resistance components with a single diode component \cite{FernandezdeCanete2013}. 

In the present work, three heart valve models are used along with the cardiovascular LPM (figure \ref{lpm_circuit}) discussed above, to study the effects of increasing degrees of aortic stenosis.

\subsubsection{Valve model 1}\label{diode_section}

The simplified heart valve model presented by \cite{Korakianitis2006a}, estimates the flow rate of blood through the $i^{th}$ valve using equation \ref{korak_q}, which is based on a typical orifice pressure drop relation \cite{Cengel2006b}. In equation \ref{korak_q}, $A_{base}$ is the base area of the inlet conduit to the valve, $\rho_l$ is the blood density, $K$ is the flow coefficient, $A_r$ is the valve area opening ratio defined as $\frac{A_{valve,open}}{A_{base}}$ and $\Delta P(t)$ is the absolute pressure gradient across the valve which is calculated using equation \ref{korak_delta_p}. 

The authors of this diode-based orifice model, designated the variable $CQ$ as the valve flow coefficient. These coefficients are set to constant values that differ for the semilunar and atrioventricular (AV) valves. For the semilunar valves, $CQ_{AO} = CQ_{PO} = 350 \left[\frac{mL}{s \cdot mmHg^{0.5}}\right]$ and for the AV valves, $CQ_{MI} = CQ_{TI} = 400 \left[\frac{mL}{s \cdot mmHg^{0.5}}\right]$. These values were tuned manually, to produce near physiological cardiovascular behaviour over a range of mitral stenosis and aortic regurgitation cases analysed in \cite{Korakianitis2006a}. 

\begin{equation}
	Q_i(t) = \sqrt{\frac{2 A_{base}^2}{\rho K}} A_{r}(t) \sqrt{\Delta P(t)} = CQ \cdot A_{r}(t) \sqrt{\Delta P(t)}
	\label{korak_q}
\end{equation}

\begin{equation}
	\Delta P(t) = 
	\begin{cases}
		P_{in}(t) - P_{ex}(t), \text{if } & P_{in} \geq P_{ex} \\
		P_{ex}(t) - P_{in}(t), \text{if } & P_{in} < P_{ex}
	\end{cases}
	\label{korak_delta_p}
\end{equation}

The valve area opening ratio $A_r$ shown in equation \ref{korak_q}, is estimated using simple diode behaviour, where $A_r$ either takes a value of 0 or 1 depending on the pressure gradients across the specific heart valve, as shown in equation \ref{korak_ar}. In equation \ref{korak_ar}, $P_{in}(t)$ $\left[ mmHg \right]$ is the instantaneous valve upstream static blood pressure and $P_{ex}(t)$ $\left[ mmHg \right]$ is the instantaneous valve downstream static blood pressure.

\begin{equation}
	A_r(t) = 
	\begin{cases}
		1, \text{if } & P_{in}(t) \geq P_{ex}(t) \\
		0, \text{if } & P_{in}(t) < P_{ex}(t)
	\end{cases}
	\label{korak_ar}
\end{equation}  

\subsubsection{Valve model 2}\label{valve_dynamics_section}
The advanced heart valve modelling methodology presented by \cite{Korakianitis2006a}, also uses equation \ref{korak_q} and accompaning model constants to calculate the pressure losses through the four heart valves. The difference between valve model 1 and 2, is in the estimation of the valvular area opening ratio $A_r$. For valve model 2, the valve opening ratio is calculated as a function of the valve cusp opening angle $\theta_{v}$, as shown in equation \ref{ar_theta}, where $\theta_{v, max}$ is the maximum opening angle of the valve cusps.

\begin{equation}
	A_r(t) = \frac{\left(1-\cos \theta_{v}(t)\right)^2}{\left( 1 - \cos \theta_{v,max} \right)^2}
	\label{ar_theta}
\end{equation} 

To find the time-dependent $\theta_{v}(t)$ $\left[ \text{rad} / \text{degrees}\right]$ for each heart valve, four differential equations governing the valve cusp motion are solved. These dynamic equations, takes into account the pressure forces acting on the valve cusps, frictional forces due to neighbouring tissue resistance and the fluid-structure interactions of the vortexes forming downstream of the valves. Therefore, to simulate the rotational cusp motion of the $i^{th}$ heart valve, equation \ref{theta_valve} is numerically solved.

\begin{equation}
	\frac{d^2 \theta_{v,i}}{dt^2} = \frac{F_{p,i}(t) - F_{f,i}(t) - F_{v,i}(t)}{I_{o,i}}
	\label{theta_valve}
\end{equation}

In the equation above, $F_{p,i}(t)$ is the pressure forces, $F_{f,i}(t)$ the tissue frictional forces and $F_{v,i}(t)$ the vortex forces acting on the $i^{th}$ heart valve at time step $t$. $I_{o,i}$ is the mass moment of inertia of the heart valve cusps. 

As discussed in the works of Korakianitis and Shi \cite{Korakianitis2006a}, \cite{Korakianitis2006} and Mynard et al. \cite{Mynard2011}, the pressure force is the dominant factor which dictate valvular motion. Therefore, in the current work only the pressure forces are considered and frictional and vortex forces ignored ($F_{f} = F_{v} = 0$). To estimate the pressure forces on the cusps of the $i^{th}$ heart valve the constitutive relationship shown in equation \ref{korak_force} is utilised. 

\begin{equation}
	F_{p,i}(t) = \left( P_{in,i}(t) - P_{ex,i}(t) \right) \cdot K_{p,i} \cos \left( \theta_{v,i}(t) \right)
	\label{korak_force}
\end{equation}

In the equation above, the valvular force coefficient $K_p$ $\left[ \frac{\text{rad}}{s^2 \cdot mmHg} \right]$ is assumed to have a constant value of 5500 for all four heart valves. This force coefficient value was determined through manual model tuning to produce valvular motion similar to that found in literature. From the above discussion it is apparent that the proposed model, does not account for all the mechanical forces acting on the AV valves such as cord tensions. Nonetheless it has been shown that the valve model 2 dynamics adequately captures the motion of the AV valves.

\subsubsection{Valve model 3}\label{new_valve_model}

The challenge associated with implementing the valve modelling methodologies discussed in sections \ref{diode_section} and \ref{valve_dynamics_section} in patient-specific LPMs, is that the flow ($CQ$) and valvular force ($K_p$) coefficients are manually tuned and, therefore, not applicable to a wide range of valve geometric and haemodynamic parameters. In the present section, valve pressure loss and motion models will be presented that are based on basic valve parameters such as valve cusp thickness, cusp height, valve opening angle and instantaneous valve flow rate. Furthermore, the pressure loss calculations will distinguish between AV and semilunar valves due to the different configurations relative to the adjacent heart chambers. The proposed model does not consider the individual valvular disease morphologies such as bicuspid aortic disease and rheumatic heart disease. Rather all diseases are assumed to only reduce the maximum valve opening angle.

\begin{figure}[h!]
	\centering
	\includegraphics[width=\textwidth]{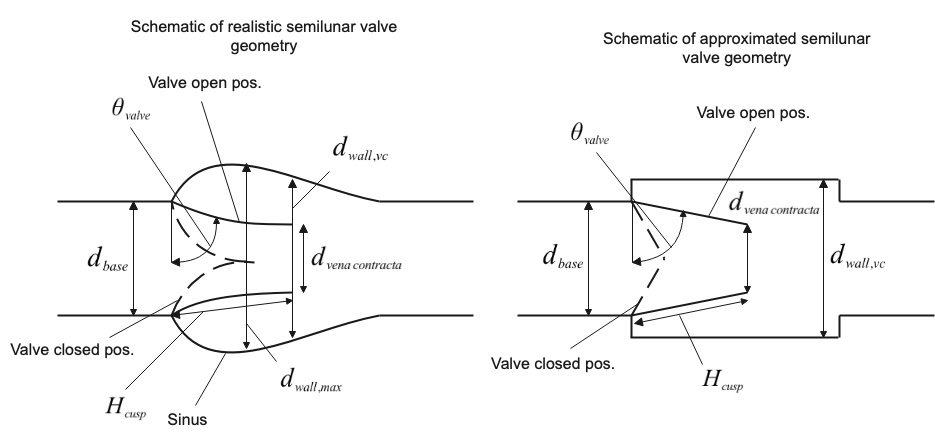}
	\caption{Schematics of semilunar valve geometries}
	\label{valves}
\end{figure}

Figure \ref{valves}, shows a schematic of a realistic semilunar heart valve geometry (left) and the approximated valve geometry (right) for the current modelling methodology. For both the AV and semilunar valves the cusps are approximated as thin straight rectangular plates rotating around hinge points located at the base of the valve.  

To simulate the valve cusp motion a similar approach is used to the one discussed in section \ref{valve_dynamics_section}. The governing dynamic differential model equations for the $i^{th}$ heart valve are shown in equations \ref{new_motion_1} and \ref{new_motion_2}, where $\omega_i$ $\left[ \frac{rad}{s} \right]$ is the valve cusp angular velocity, $\bar{L}_i$ $\left[ m \right]$ is the distance from the base of the cusp to its cusp centre of gravity and $A_{base} = \pi d_{base}^2/4$ is the closed valve flow area. The factor of $133.\dot{3}$ is used to convert from SI units to $\left[ \frac{\text{rad}}{s^2 \cdot mmHg} \right]$.

\begin{equation}
	\frac{d \theta_{v,i}}{dt} = \omega_{i}(t)
	\label{new_motion_1}
\end{equation}

\begin{equation}
	I_{o,i} \frac{d \omega_i}{dt} = \left( \left[ P_{in,i}(t) - P_{ex,i}(t) \right] \cdot A_{base,i} \cdot \bar{L}_i \cdot \cos \theta_{v,i} \right) \cdot 133.\dot{3}
	\label{new_motion_2}
\end{equation}

The mass moment of inertias for the cusps are calculated using equation \ref{new_motion_3} which is based on the standard expression for mass moment of inertia of a rectangular plate rotating about its edge \cite{Meriam2003}. The mass of the $i^{th}$ valve is calculated as $m_{valve,i} = \pi d_{base} H_{cusp,i} t_{cusp,i} \rho_{cusp}$, where $H_{cusp,i}$ $\left[ m \right]$ is the height of the valve cusps when fully open (see figure \ref{valves}), $\rho_{cusp}$ $\left[ \frac{kg}{m^3} \right]$ is the density of the cusp material and $t_{cusp,i}$ $\left[ m \right]$ is the thickness of the valve cusps. For stenosed valves it was assumed that the cusp thicknesses remain unchanged.

\begin{equation}
	I_{o,i} = \frac{m_{valve,i}}{12} \cdot \left( H_{cusp,i}^2 - t_{cusp,i}^2 \right) + m_{valve,i} \cdot \bar{L}_i^2
	\label{new_motion_3}
\end{equation}

If one studies the geometry of the heart valve shown in figure \ref{valves} from a pure fluid dynamics point-of-view, the valve geometry and accompanying flow losses are more akin to a gradual contraction or a nozzle with variable wall angles, rather than an variable area orifice plate. Therefore, the proposed model uses the calculated valve cusp angles, cusp heights and valvular flow rates to determine the time-dependent flow coefficients of each valve based on a simplified nozzle/contraction geometry. Below the calculation procedures for the AV and semilunar valves will be discussed.

\begin{figure}[h!]
	\centering
	\includegraphics[width=\textwidth]{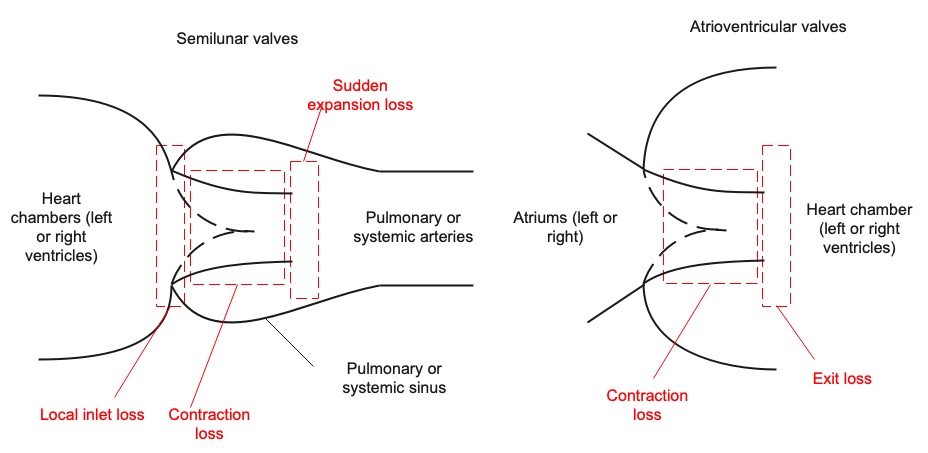}
	\caption{AV and semilunar valve losses}
	\label{valve_losses}
\end{figure}

Blood is forced into the semilunar valves by the corresponding ventricular contraction and through the valve body and into the downstream sinus. For the semilunar valves the total pressure loss comprises of a local inlet loss due to the blood entering the valve body, a contraction loss due to frictional and geometrical effects as the blood travels through the nozzle-shaped geometry and a sudden expansion loss due to blood exiting the valve body and entering the larger flow area of the sinus. The location of these losses can be seen in figure \ref{valve_losses}. Seeing as the nozzle formed by the valve cusps protrudes into the sinus, the difference in flow area experienced by the blood as it exits the valve, is not $d_{vena,contracta}\rightarrow d_{wall,max}$ but rather $d_{vena,contracta} \rightarrow d_{wall,vc}$, where $d_{vena,contracta} = d_{base} \cdot \sqrt{A_r}$. The typical ratio of base diameter $d_{base}$ to maximum sinus diameter $d_{wall,max}$ for semilunar valves are 1.46 \cite{Swanson1974}, \cite{Sands1969}, in the present work the sinus diameter at the exit of the nozzle is assumed to be $1.23 \cdot d_{base}$ as seen in figure \ref{valves}.

To calculate the flow rate through the semilunar valves using the proposed valve model, equation \ref{new_q_valve} is used, where the $133.\dot{3}$ and $10^6$ factors are again used to convert from SI to medical units $\left( mL/s \right)$. 

\begin{equation}
	Q_i(t) = \left( \sqrt{\frac{2 A_{base}^2 A_r^2}{\rho_l K_{t}(t)}\Delta P(t) \cdot 133.\dot{3} } \right)\cdot 10^6
	\label{new_q_valve}
\end{equation}

In the equation above $K_t$ is the total flow coefficient for the semilunar valves and consists of $K_{c,SL}$ the local contraction loss coefficient, $K_{f,SL}$ the semilunar valve friction loss coefficient and $K_{se,SL}$ the semilunar valve sudden expansion coefficient. The contraction loss coefficient is calculated as shown in equation \ref{k_contract} \cite{Levin1970}, \cite{Rennels2012}.

\begin{equation}
	K_{c,SL}(t) = 0.0696\sin(\alpha_i(t))\cdot(1 - \beta_i(t)^5)(\lambda_i(t)^2) + (\lambda_i(t) - 1)^2
	\label{k_contract}
\end{equation}

In equation \ref{k_contract}, $\alpha_i(t) = \frac{\pi}{2} - \theta_{v,i}(t)$ and $\beta_i(t) = \sqrt{A_r(t)}$ for the $i_{th}$ valve. The contraction ratio is calculated as $\lambda_i(t) = 1 + 0.622 \cdot \left( \frac{\alpha_i(t)}{180}  \right)^{4/5} \cdot \left( 1 - 0.215 \beta_i(t)^2 - 0.785 \beta_i(t)^5 \right)$. The frictional loss coefficient is calculated using equation \ref{k_friction}.

\begin{equation}
	K_{f,SL}(t) = \frac{f_i \left( 1 - \beta_i(t)^4 \right)}{8 \sin\left( \frac{\alpha_i(t)}{2} \right)} + \beta_i(t)^4 \frac{f_i(t) H_{cusp,i}}{0.5\left(d_{base,i}(t) + d_{vena,contracta}(t)\right)}
	\label{k_friction}
\end{equation}

The friction factor $f_i$ per semilunar valve is calculated using the well-known smooth surface Colebrook-White equation \cite{Cengel2006b} shown equation \ref{colebrook}. The Reynolds number used in the calculation of the friction factor is determined using equation \ref{Re_number}, and is based on the velocity at the vena contracta of the semilunar valve.

\begin{equation}
	\frac{1}{\sqrt{f_i}} = -2 \log_{10} \left( \frac{2.51}{Re_i \sqrt{f_i}} \right)
	\label{colebrook}
\end{equation}

\begin{equation}
	Re_i = \frac{\rho_l v_i(t) d_{vena,contracta}(t)}{\mu_l}, \ \text{where} \ v_i(t) = \frac{Q_i(t)}{A_{base} A_r(t)}
	\label{Re_number}
\end{equation}

The sudden expansion loss coefficient for the semilunar valves is calculated using equation \ref{se_loss} \cite{Rennels2012}, where $\beta_{2,i}(t) = \frac{d_{vena,contracta}(t)}{d_{wall,vc}}$.  

\begin{equation}
	K_{se,SL}(t) = \left( 1 - \beta_{2,i}(t)^2 \right)^2
	\label{se_loss}
\end{equation}

For the AV valves, blood is ejected from the atriums through the valves into the ventricles. Seeing as the ventricles act as blood reservoirs and have larger volumes than the upstream atriums, the AV valve geometries are approximated as smooth nozzles discharging into large static fluid volumes, as seen in figure \ref{valve_losses}. Similar to the semilunar valves, the flow rates through the AV valves are calculated using equation \ref{new_q_valve}. The total loss coefficient for the AV valves are calculated using equation \ref{k_semilunar}, where the AV valve frictional loss coefficient $K_{f,AV}$ is calculated using the same approach as for the semilunar valves. It should be noted that AV valves typically have $\theta_{v,max}$ values larger than $75^0$, but for the purpose of the present study the values proposed by Korakianitis and Shi will be used.

\begin{equation}
	K_t(t) = K_{f,AV}(t) + 1
	\label{k_semilunar}
\end{equation}

The physiological parameters used in valve model 3 such as cup thicknesses, maximum valve opening angles and valve base diameters were taken from literature for an average human male, and the values can be seen in table \ref{valve_data}.

\begin{table} [h!]
\centering
	\caption{Valve model 3 physiological parameters}
	\begin{tabular} {c c c c}
		\hline
		Parameter & Value & Units & Reference \\
		\hline
		\emph{Aortic valve} \\
		base diameter, $d_{base,AO}$ & 24.7 & $\left[ mm \right]$ & \cite{Westaby1984} \\
		cusp height, $H_{cusp,AO}$ & 17.5 & $\left[ mm \right]$  & \cite{Swanson1974}, \cite{Sands1969}	 \\
		cusp thickness, $t_{cusp,AO}$ & 0.61 & $\left[ mm \right]$ & \cite{Stradins2004}	\\ 
		sinus diameter, $d_{wall,vc,AO}$ & 30.4 & $\left[ mm \right]$ & \cite{Swanson1974}, \cite{Sands1969}	 \\ 
		maximum opening angle, $\theta_{v,max,AO}$ & 75 & $\left[ \text{degrees} \right]$ & \cite{Korakianitis2006a} \\
		minimum opening angle, $\theta_{v,min,AO}$ & 5 & $\left[ \text{degrees} \right]$ & \cite{Korakianitis2006a} \\
		\emph{Pulmonary valve} \\
		base diameter, $d_{base,PO}$ & 25 & $\left[ mm \right]$ & \cite{Westaby1984} \\
		cusp height, $H_{cusp,PO}$ & 17.6 & $\left[ mm \right]$  & $1.42\frac{d_{base,PO}}{2}$	 \\
		cusp thickness, $t_{cusp,PO}$ & 0.4 & $\left[ mm \right]$ & \cite{Stradins2004}	\\ 
		sinus diameter, $d_{wall,vc,PO}$ & 30.6 & $\left[ mm \right]$ & $1.23d_{base,PO}$ \\ 
		maximum opening angle, $\theta_{v,max,PO}$ & 75 & $\left[ \text{degrees} \right]$ & \cite{Korakianitis2006a} \\
		minimum opening angle, $\theta_{v,min,PO}$ & 5 & $\left[ \text{degrees} \right]$ & \cite{Korakianitis2006a} \\
		\emph{Mitral valve} \\
		base diameter, $d_{base,MI}$ & 27.0 & $\left[ mm \right]$ & \cite{Dwivedi2014} \\
		cusp height, $H_{cusp,MI}$ & 19.17 & $\left[ mm \right]$  & $1.42\frac{d_{base,MI}}{2}$	 \\
		cusp thickness, $t_{cusp,MI}$ & 1.3 & $\left[ mm \right]$ & \cite{Stradins2004}	\\ 
		maximum opening angle, $\theta_{v,max,MI}$ & 75 & $\left[ \text{degrees} \right]$ & \cite{Korakianitis2006a} \\
		minimum opening angle, $\theta_{v,min,MI}$ & 5 & $\left[ \text{degrees} \right]$ & \cite{Korakianitis2006a} \\
		\emph{Tricuspid valve} \\
		base diameter, $d_{base,TI}$ & 28.0 & $\left[ mm \right]$ & \cite{Dwivedi2014} \\
		cusp height, $H_{cusp,TI}$ & 19.9 & $\left[ mm \right]$  & $1.42\frac{d_{base,TI}}{2}$	 \\
		cusp thickness, $t_{cusp,TI}$ & 0.9 & $\left[ mm \right]$ & \cite{Stradins2004}	\\ 
		maximum opening angle, $\theta_{v,max,TI}$ & 75 & $\left[ \text{degrees} \right]$ & \cite{Korakianitis2006a} \\
		minimum opening angle, $\theta_{v,min,TI}$ & 5 & $\left[ \text{degrees} \right]$ & \cite{Korakianitis2006a} \\
		\hline
	\end{tabular}
	\label{valve_data}
\end{table}

For valve model 3, the blood viscosity, blood density and cusp material density used are $\mu_l = 0.0035$ $\left[ Pa \cdot s \right]$ \cite{Yan2021a}, $\rho_l = 995$ $\left[ \frac{kg}{m^3} \right]$ \cite{Vitello2015}, $\rho_{cusp} = 1060$ $\left[ \frac{kg}{m^3} \right]$ \cite{Yan2021a}.

\subsection{Case studies}

In the current work, the valve models presented in sections \ref{diode_section}, \ref{valve_dynamics_section}, \ref{new_valve_model} are applied to simulate the local and global haemodynamics of a typical human male cardiovascular system. The results generated using the three valve models are compared for different aortic stenosis case studies. Each successive case study corresponds to a reduction in maximum opening flow area of the valve. For valve model 1, the area opening ratio $A_r$ is simply limited to a specified value and for valve models 2 and 3 the maximum opening angles are limited $\theta_{v,max,AO}$. Table \ref{cases}, contains the limits applied to the aortic valve for the 5 case studies.

\begin{table}[h!]
	\centering
	\caption{Aortic stenosis case studies}
	\begin{tabular} {c c c c c}
		\hline
		Case \# & Description & Valve area $\left[ cm^2 \right]$ &  $A_{r,max,AO}$ &  $\theta_{v,max,AO}$ $\left[ \text{degrees} \right]$ \\
		\hline
		1 & None & 4.81 & 1 & 74.9 \\
		2 & Moderate & 1.63 & 0.333 & 55.1 \\
		3 & Severe & 1.06  & 0.222 & 49.4 \\
		4 & Very severe & 0.64  & 0.1333 & 43.2 \\
		5 & Critical & 0.43  & 0.0899 & 38.8 \\
		\hline
	\end{tabular}
	\label{cases}
\end{table}

\subsection{Numerics}
To simulate the dynamics of the cardiovascular system, the differential-algebraic equations (DAEs) for the time-varying elastance model (section \ref{heart_section}), vasculature dynamics (section \ref{vasculature_section}) and the heart valve dynamics (section \ref{heart_valve_section}) are solved simultaneously using an ODE system of equations integrator. For the current work the Tsitouras 5/4 Runge-Kutta method \cite{Tsitouras2011} is used with the relative and absolute tolerances set to $1\cdot10^{-4}$ and $1\cdot10^{-6}$ respectively. The maximum integrator iteration count of $10^6$ was also specified.

To numerically integrate and solve the mentioned DAEs, certain constraints and conditions had to be incorporated such as valve motion limits, implicit equation solving and initial conditions. To incorporate the discontinuities arising from the valve motion limits the following condition is included in the simulation procedure of each valve

\begin{equation}
	\theta_{v} =
	\begin{cases}
		\theta_{v} = \theta_{v,max}, \ \omega = 0.0 & \text{ if } \theta_{v} \geq \theta_{v,max} \\
		\theta_{v} = \theta_{v,min}, \ \omega = 0.0 & \text{ if } \theta_{v} \leq \theta_{v,min} \\
		\theta_{v} & \text{ if } \theta_{v,min} < \theta_{v} < \theta_{v,max}
	\end{cases}
\end{equation}

To find the friction factor for each valve per time step requires the solution of the implicit Colebrook-White equation (see equation \ref{colebrook}). In the DAE integrator solver loop of the cardiovascular system, a nested Newton-Raphson solver is used to find the friction factors for a given valve, flow rate, and valve opening fraction. 

To solve the DAEs, the initial conditions are required. For the current model, initially the semilunar valves were set to fully open and the AV valves to fully closed. Furthermore, the initial volume of the left and right ventricles are set to 800 and 500 $mL$ respectively and the initial volumes of both atriums to 2.5 $mL$. For the remaining pressure and flow rate dependent variables the initial conditions are set to 1 $mmHg$ and 0 $\frac{mL}{s}$ respectively. The initial valve cusp angular velocities are set to $0.0$ $\frac{\text{rad}}{s}$.

\section{Results and discussions}

To demonstrate that the proposed LPM, along with the three heart valve modelling extensions, can reproduce typical physiological haemodynamic parameters the results for case study 1 was compared to data from literature as seen in table \ref{comparison_to_physio}. Note that M1, M2 and M3 corresponds to the LPM implemented with valve models 1, 2 and 3 respectively.

\begin{table}[h!]
	\centering
	\caption{Comparison between LPMs and typical physiological haemodynamic parameters}
	\begin{tabular} {c c c c c}
		\hline
		Parameters & M1 & M2 & M3 & Physiology \\
		\hline
		LV Stroke volume, $mL$ & 82.7 & 73.9 & 73.1 & 51-111 \cite{Rosalia2021} \\
		Cardiac output, $L/min$ & 4.9 & 4.4 & 4.4 & 4.0-8.0 \cite{LIDCO2022} \\
		\emph{Left ventricle} \\
		Systolic pressure, $mmHg$ & 123.9 & 115.8 & 115.6 & 90-140 \cite{Albanese2016} \\
		Diastolic pressure, $mmHg$ & 7.5 & 7.3 & 7.0 & 4-12 \cite{Albanese2016} \\
		Systolic volume, $mL$ & 53.5 & 50.4 & 50.4 & 37-57 \cite{Rosalia2021} \\
		Diastolic volume, $mL$ & 137.2 & 140 & 140.8 & 121-163 \cite{Rosalia2021} \\
		\emph{Right ventricle} \\
		Systolic pressure, $mmHg$ & 29.7 & 28.9 & 28.5 & 15-30 \cite{Rosalia2021} \\
		Diastolic pressure, $mmHg$ & 6.4 & 6.7 & 6.3 & 2-8 \cite{Rosalia2021} \\
		Systolic volume, $mL$ & 37 & 36.7 & 36.7 & 36-84 \cite{Rosalia2021} \\
		Diastolic volume, $mL$ & 122 & 126.3 & 124.9 & 121-167 \cite{Rosalia2021} \\
		\emph{Systemic arterial pressures} \\
		Systolic, $mmHg$ & 123.6 & 115.6 & 115.4 & 90-140 \cite{Albanese2016} \\
		Diastolic, $mmHg$ & 84.2 & 76.4 & 75.6 & 60-90 \cite{Albanese2016} \\
		\emph{Pulmonary arterial pressures} \\
		Systolic, $mmHg$ & 28.9 & 28.3 & 28.3 & 15-30 \cite{Rosalia2021} \\
		Diastolic, $mmHg$ & 15.9 & 16.6 & 16.7 & 5-16 \cite{Albanese2016} \\
		\hline
	\end{tabular}
	\label{comparison_to_physio}
\end{table}

The results in table \ref{comparison_to_physio} shows that the three models can capture typical physiological parameters of the human cardiovascular system. It is seen that M2 and M3 produce very similar results for case study 1, whereas M1 generates results with a higher cardiac output and higher blood pressures across the network. These differences between models are due to the different valvular pressure drops calculated by each approach and are reflected in table \ref{dp_case_1}. 

\begin{table}[h!]
\caption{Peak pressure drops, $\text{max}\left( \Delta P \right)$, over valves for case study 1}
\centering
	\begin{tabular}{c c c c}
	\hline
		Valve & M1 & M2 & M3 \\
	\hline
	AO & 9 & 11.8 & 9.5 \\
	PO & 4.7 & 5.2 & 2.9 \\
	MI & 4.2 & 6 & 6.9 \\
	TI & 2.2 & 4.2 & 4.6 \\
	\hline
	\end{tabular}
	\label{dp_case_1}
\end{table}

Figure \ref{fig_areas} below shows the mitral and aortic valve area opening ratios ($A_r$) as a function time for the three valve models and selected case studies (1,2,4). The diode-like behaviour of the M1 valves can be seen in the results as the valves abruptly opens and closes, with no inertia. For case studies 1 and 2, M2 and M3 produces very similar results. It is observed that the inclusion of the valve motion effects results in the valves opening later and closing more slowly compared to the M1 results. In general, it can be seen that the developed fundamental valve motion model (valve model 3) which uses geometric and fluid property parameters rather than empirically tuned values can recreate expected valve motion \cite{Korakianitis2006}.
 
\begin{figure*}[h!]
	\centering
	\includegraphics[width=\textwidth]{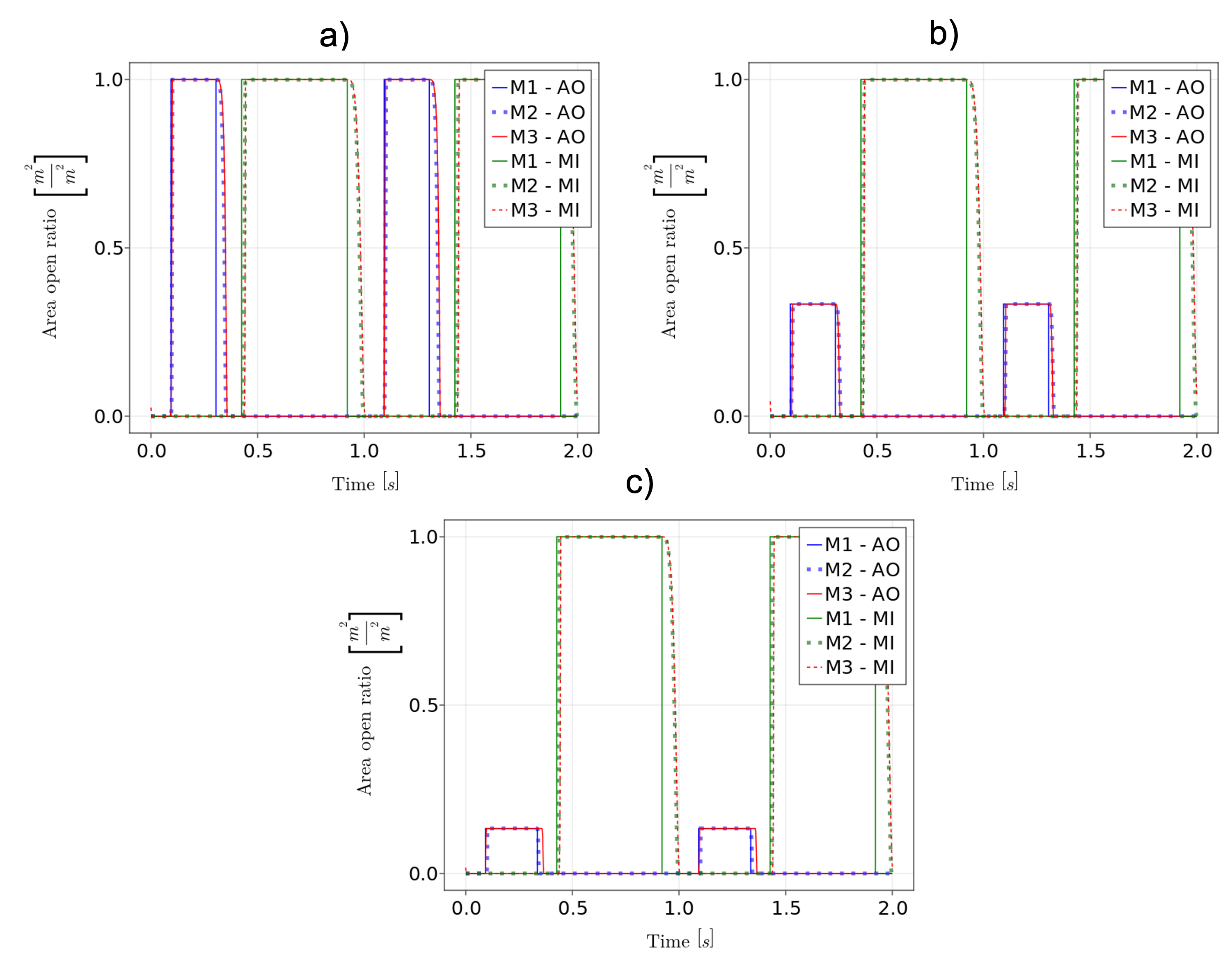}
\caption{Time-dependent aortic and mitral valve area opening ratios (a) Case study 1, (b) Case study 2, (c) Case study 4}
\label{fig_areas}
\end{figure*}

\begin{figure*}[h!]
	\centering
	\includegraphics[width=\textwidth]{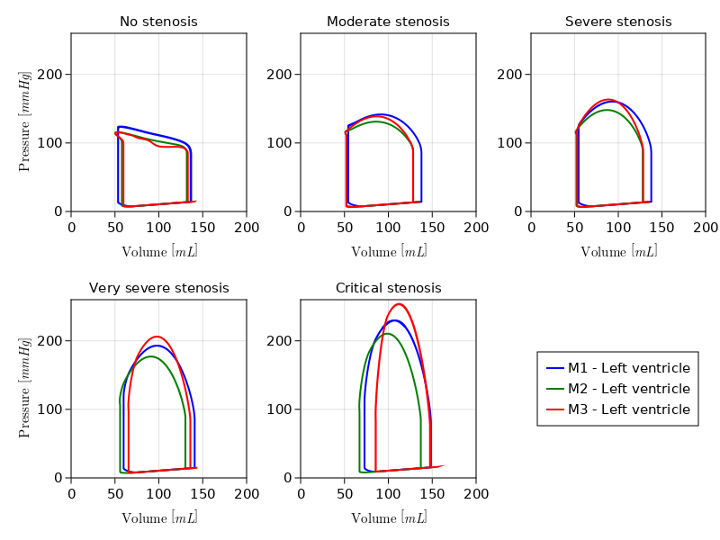}
	\caption{Left ventricle pressure-volume loops for increasing degrees of aortic stenosis calculated using valve models 1,2,3}
	\label{fig_pv}
\end{figure*}

To investigate the effect that the different valve modelling methodologies have on global haemodynamic parameters for increasing degrees of aortic stenosis, the left ventricle pressure-volume (PV) loops for the case studies are plotted in figure \ref{fig_pv}. Comparing M1 and M2 results, it is observed that the left ventricular systolic pressures (LVSPs) predicted by M2 are lower than the values generated using M1. This is in agreement with the observations by Korakianitis and Shi \cite{Korakianitis2006a}. The percentage differences between M1 and M2 LVSPs over the range of stenosis cases analysed stay relatively constant at 7.6\%. M1 predicts on-average 9\% higher stroke volumes compared to the values predicted by M2. Therefore, M2 predicts lower cardiac output, which highlights the importance of including valve motion effects seeing as cardiac output is a critical clinical parameter \cite{Rosalia2021}. A larger range of LVSP differences are observed when comparing the PV loops of M1 and M3. For case study 1, the M3 predicted LVSP is approximately 8.8\% lower than the M1 predicted LVSP and for case study 5, the M3 predicted LVSP is 10.4\% higher than the value predicted by M1. This shows a changing relationship in LVSP differences driven by the valvular pressure drop calculation methodology. Similarly, a larger variation in stroke volume differences between M1 and M3 are observed. For case study 1, the stroke volume predicted by M3 is 11.6\% lower than the value predicted by M1 and for case study 5, the M3 predicted stroke volume is 18.2\% lower than the M1 value. The LVSP and stroke volume trends observed in figure \ref{fig_pv}, indicates that the M3 modelling approach predicts higher pressure drops (ventricular afterload) in the stenosed valve compared to the M1 and M2 models.

\begin{figure*}[h!]
	\centering
	\includegraphics[width=\textwidth]{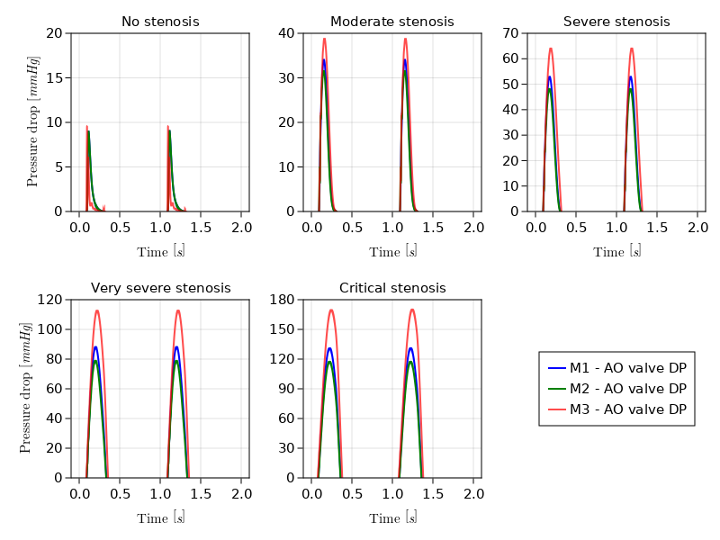}
	\caption{Pressure drops over aortic valve for different degrees of stenosis}
	\label{fig_dp}
\end{figure*}

Figure \ref{fig_dp} shows the aortic valvular pressure drops calculated using the three different modelling approaches over the range of analysed aortic stenosis cases. The results show for the different cases, that M1 and M2 valve modelling approaches calculates similar pressure drops, with an average peak pressure drop difference between the modelling approaches of approximately 9\%. The M3 valve modelling approach, on the other hand, calculates significantly higher pressure drops for the stenosed cases. The peak pressure drops calculated by M2 and M3 for case study 1 are 12 and 9.5 $mmHg$ respectively. For case study 4, 79.1 $mmHg$ (M2) and 113.1 $mmHg$ (M3) and for case study 5, 117 $mmHg$ (M2) and 170.1 $mmHg$ (M3). Handke et al. \cite{Handke2003} performed in-vivo transesophageal 3D echocardiography to investigate aortic valve dynamics. The authors reported for mild ($A_r = 0.5$) to severe ($A_r = 0.15$) aortic stenosis the range of peak valvular pressure drops are 40-130 $mmHg$. Using the data generated in the present work, the peak pressure drops for M2 and M3 were interpolated to the $A_r$ values mentioned by Handke et al. For $A_r = 0.5$, the M2 and M3 values are 26.68 $mmHg$ and 38.5 $mmHg$ respectively. For $A_r = 0.15$, the M2 and M3 values are 73.1 $mmHg$ and 104 $mmHg$ respectively. Messika-Zeitoun and Lloyd \cite{Messika-Zeitoun2018}, reported that for stenosed aortic valve flow areas of 1.5 $cm^2$ and 1 $cm^2$ the mean transvalvular gradients are 20 $mmHg$ and 40 $mmHg$ respectively. For similar flow areas the M3 approach generates aortic valve gradients of 21.8 $mmHg$ and 42.5 $mmHg$, where the M2 approach generates gradients of 16.2 $mmHg$ (M1 - 17.7 $mmHg$) and 30.0 $mmHg$ (M1 - 33 $mmHg$). These results show that the M3 modelling approach can capture the experimentally measured range of typical aortic valve pressure drops more accurately than the M1 and M2 approaches. 

\begin{figure*}[h!]
	\centering
	\includegraphics[width=\textwidth]{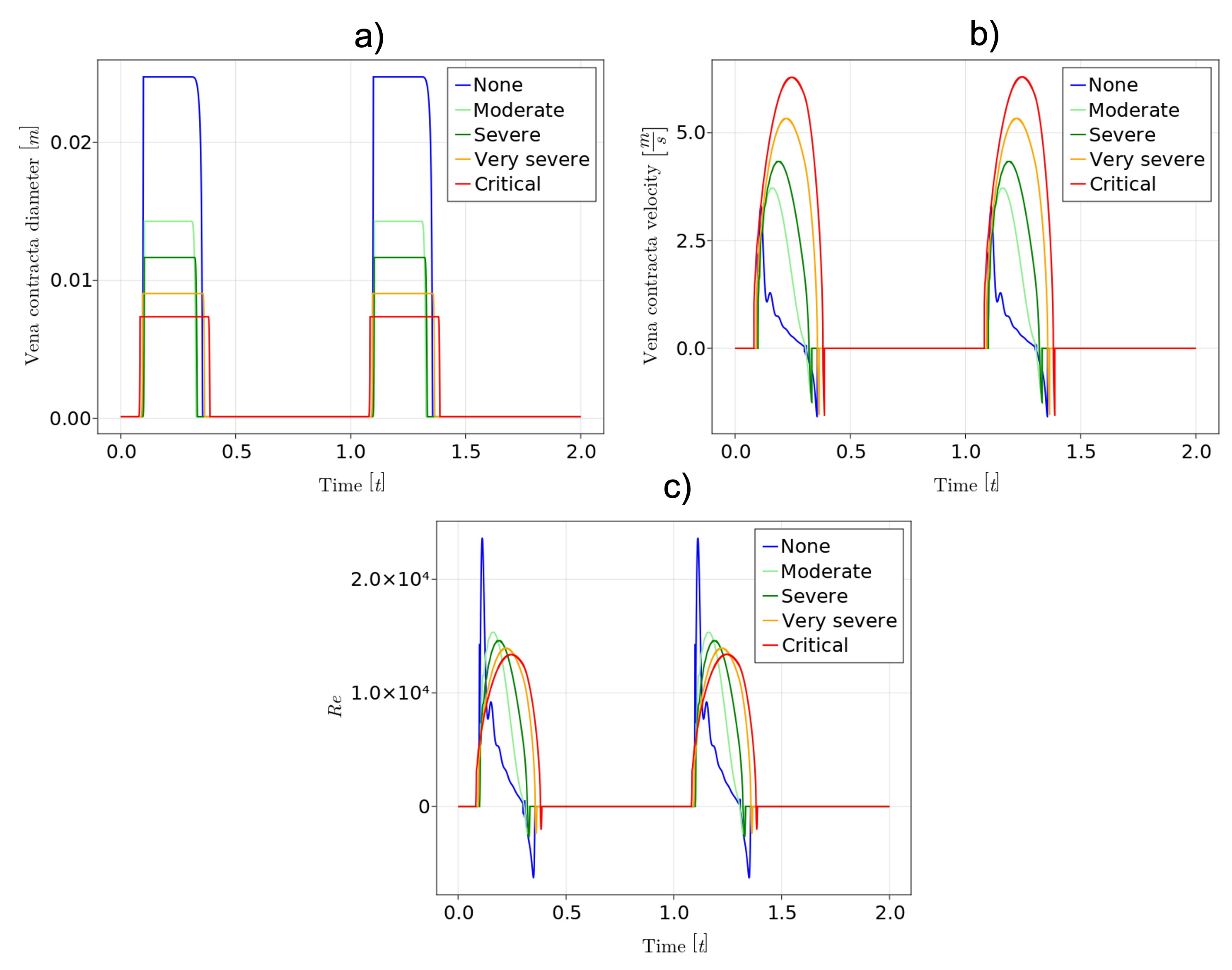}
	\caption{Time-dependent (a) valve opening diameters, (b) vena contracta velocities and (c) Reynolds numbers for valve model 3}
	\label{fig_pars}
\end{figure*}

To understand why the M3 modelling approach is predicting significantly higher pressure drops over the range of aortic stenosis cases, the time-dependent valve opening diameters, vena contracta velocities and Reynolds numbers are plotted in figure \ref{fig_pars}. The opening diameters of the aortic valve are calculated as $d_{vena,contracta} = 2\sqrt{\left(A_r \cdot A_{base}\right)/\pi}$. In figure \ref{fig_pars} (a,b), it is seen as the maximum opening diameter of the vena contracta reduces due to stenosis, the peak and duration of the high flow velocities in the valve body increases. For the case with no stenosis the initial peak vena contracta velocity is approximately 2.5 $\frac{m}{s}$, but this velocity corresponds to the time step at which the valve starts to open. The peak vena contracta velocity for case study 1 (no stenosis) when transvalvular flow is established is approximately 1.25 $\frac{m}{s}$ as seen in figures \ref{fig_pars} (a) and (b). For the severe and very severe cases the peak velocities are above 4.0 $\frac{m}{s}$. These values agree well with physiological values found in literature. Garcia et al. \cite{Butz2017}, stated that for healthy aortic valves the peak vena contracta flow velocity is between 1-1.5 $\frac{m}{s}$ for males between 21 to 59 years of age. Messika-Zeitoun and Lloyd \cite{Messika-Zeitoun2018}, stated velocities of  $>4$ $\frac{m}{s}$ for cases where the valve areas are below $1$ $cm^2$. 

The increase in duration of high velocities in the valve, results in sustained high flow Reynolds numbers in the valve as seen in figure \ref{fig_pars} (c). From the results, it is seen that the blood flow Reynolds numbers in the valve changes significantly over the ejection period as well as over the various degrees of stenosis. Therefore, based on equation \ref{k_friction} the flow coefficient will not remain constant as proposed by M1 and M2. Studying the Reynolds number results further, it is seen that the blood flow is turbulent for most of the ejection time duration, therefore, the assumption of using the turbulent smooth tube friction factor model of Colebrook-White is vindicated.

\begin{figure*}[h!]
	\centering
	\includegraphics[width=\textwidth]{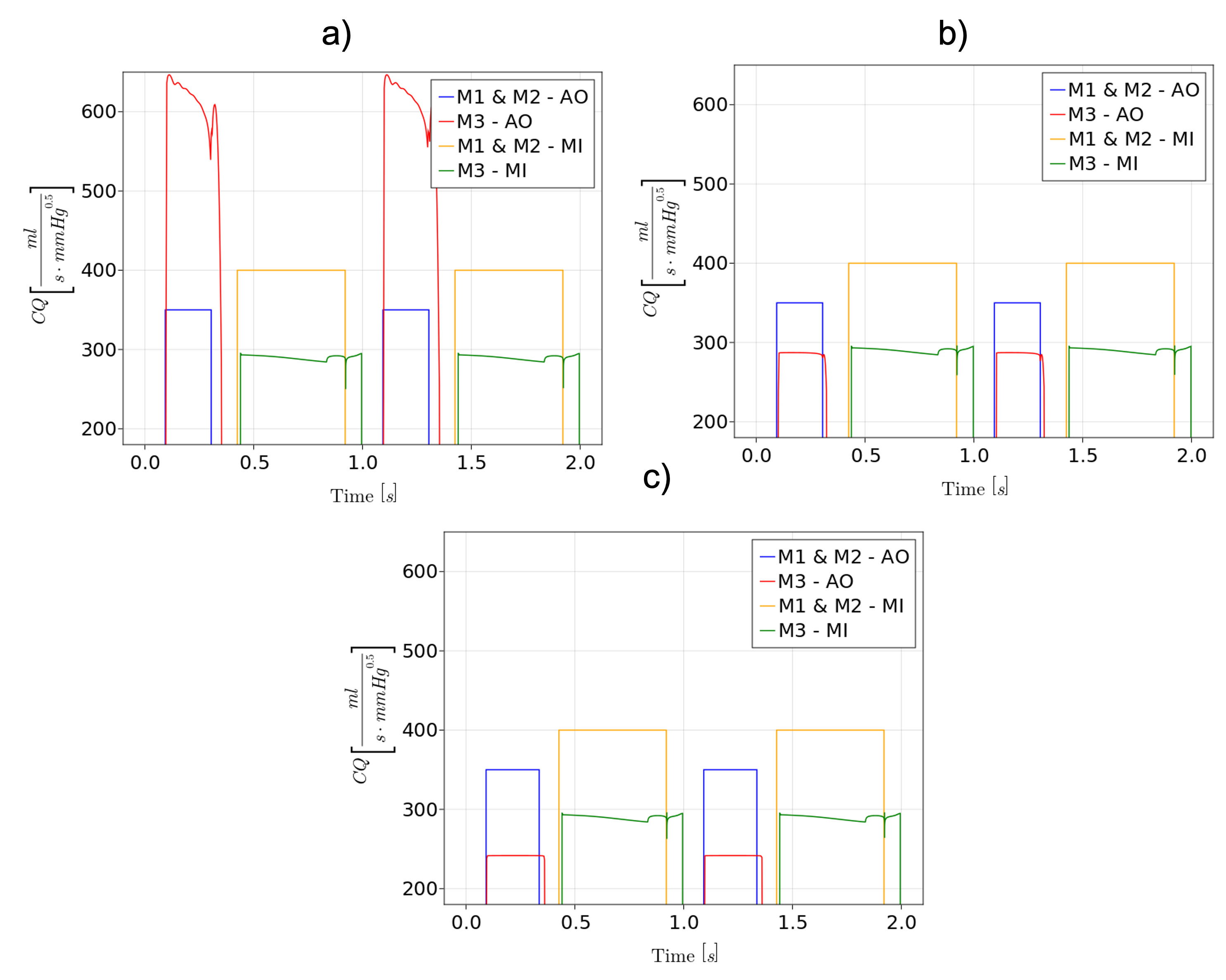}
	\caption{Time-dependent flow coefficients for models M1, M2 and M3. (a) No-stenosis, (b) Moderate stenosis and (c) Very severe stenosis.}
	\label{fig_cq}
\end{figure*}

To show the effect of taking into account the variation in aortic valve Reynolds number for the range of analysed cases, the time-dependent flow coefficients for models M1, M2 and M3 are plotted in figure \ref{fig_cq}. The figures show that the flow coefficient values for the semilunar and AV valves remain constant for M1 and M2, as discussed in section \ref{diode_section}. The flow coefficient values calculated by the M3 approach for the mitral valve is seen to be lower than the constant value proposed by \cite{Korakianitis2006a}, which results in M3 calculating higher pressure drops through the mitral valve when compared to the values of M2 (table \ref{dp_case_1}). Studying the aortic valve data, it is observed for case study 1, that the calculated flow coefficients by M3 are significantly larger than the constant value used by M1 and M2. This explains why in table \ref{dp_case_1}, the M3 model predicts a lower aortic valvular pressure drop compared to M2. In figure \ref{fig_cq} (b), for the moderate stenosis case we seen that the flow coefficients calculated by the M3 approach, are lower compared to the constant value used in M1 and M2. The lower flow coefficient values result in larger pressure drops over the valve during valvular ejection. For the very severe aortic stenosis case it is observed that the flow coefficients are lower than in the moderate stenosis case. This reduction in flow coefficients as the degree of stenosis is increased, is due to the higher sustained blood flow Reynolds numbers in the valve which leads to larger frictional and geometrical contraction losses. 

The approximate maximum and minimum flow coefficients calculated by the M3 modelling approach are 650 and 150 respectively, over the range of case studies. This gives an average flow coefficient value of 400 which is close to the tuned value proposed by Korakianitis and Shi \cite{Korakianitis2006a} (350 and 400). The advantage of using the M2 approach, is that it solves significantly faster than the M3 approach. The M1 and M2 modelling approaches, simulates 10 s, in approximately 15 s, whereas M3 takes 10-15 minutes.

\section{Conclusions}
 The results in the present work shows that the newly proposed valve modelling approach, predicts higher pressure drops at more severe degrees of aortic stenosis when compared to the results of the models from literature. This is due to the fact that the models from literature are insensitive to changes in blood flow Reynolds number through the valve. Comparing the calculated ranges of pressure drops over stenosed valves with experimental data from literature, shows that the proposed model captures the expected pressure drops more accurately compared to the models from literature. 
\section*{References}
\bibliographystyle{elsarticle-num}
\bibliography{Cardiovascularmodelling.bib}

\end{document}